\documentclass[aps,prb,twocolumn,showpacs,showkeys]{revtex4}%{article %a4paper %book twocolumn
\usepackage{graphicx, draftcopy, amsmath, bm, times}
\begin{document}
\title{From Mott insulator to band insulator: a DMFT study.}
%\author{A.~Fuhrmann, D.~Heilmann and H.~Monien}
\author{Andreas \surname{Fuhrmann}} 
\author{David \surname{Heilmann}}
\author{Hartmut \surname{Monien}} 
\affiliation{
  Physikalisches Institut, Universit\"at Bonn, Nu{\ss}allee 12, 53115 Bonn, Germany
}
\date{\today} \received{\today}
%\begin{landscape}
\begin{abstract} 
  The question if a Mott insulator and a band insulator are
  fundamentally different %is still open. 
 has been the matter of intensive research recently.
Here we consider a simple
  model which allows by tuning one parameter to go continously from a
  Mott insulator to band insulator. The model consists of two Hubbard
  systems connected by single particle hopping. The Hubbard
  Hamiltonian is solved by the Dynamical Mean-Field theory using %the
  Quantum Monte Carlo to solve the resulting the quantum impurity
  problem. The quasiparticle spectral function is calculated. Here we
  focus on the optical conductivity and in particular on the Drude
  weight which %clearly differentiates between the Mott insulator and
  % band insulator transition and 
  can be experimentally measured. %We
%  discuss in detail the behavior of the Drude weight close to the
%  finite temperature phase transition. 
  From our calculation we
  conclude that there is a continous crossover from the band insulator
  to the Mott insulator phase at finite temperature.
\end{abstract}
\pacs{71.10.Fd, 71.27.+a, 71.30.+h, 72.15.-v}%, 72.20.-i}
\keywords{strongly correlated electrons, metal-to-insulator transition,
Dynamical Mean-Field theory}
\maketitle
\section{Introduction}
Recently the question whether a Mott insulator and a band insulator are
fundamentally different has been raised \cite{Essler:2001,Dzyaloshinskii:2003,Essler:2005,Konik:2005,Berthod:2006,Rosch:2006,Stanescu:2006,Yang:2006}. % {\bf See
%  Rosch preprint ref. 1-6}.  
To study this question, we consider the
simplest model which allows, by tuning one parameter, to obtain a Mott
insulator as well as a band insulator phase. The model consists of two
Hubbard systems with strong on-site Coulomb repulsion which are
connected by single particle hopping.  This model can be viewed as a
model for two planes of strongly correlated electrons on a square
lattice with on-site interaction and a hopping connecting corresponding sites of 
%each neighbor site in 
the two planes.  At half filling with no interaction
the metal-to-band insulator transition is driven by increasing the hopping
between the two subsystems, i.~e., the splitting of the bonding and
antibonding bands produces a gap. In the case of %with 
no hopping between
the planes, a Mott %insulator 
transition is driven by increasing the
on-site Coulomb repulsion which localizes the electrons by suppressing
the hopping between different sites. The overall scale of the problem
is set by the hopping matrix element $t$ within the plane (which we set to
unity) %one) 
and the parameter differentiating between the Mott and band
insulator is the ratio of the hopping matrix element between the planes,
$t_\perp$, and the on-site Coulomb repulsion $U$. The approximation
used in this work consists of letting the coordination number of the
sites in each plane (4) to go to infinity.  This model has been
studied with Dynamical Mean-Field theory (DMFT) approximation using
%the 
iterated perturbation theory (IPT) at zero temperature
\cite{Monien:1997,Moeller:1999}.  We study this model at finite
temperature using QMC \cite{Hirsch1983,HirschFye1986} as impurity solver. Also, first
successful attempts to apply a more demanding continuous-time QMC algorithm to a simplified 
two-impurity problem already exist\cite{Savkin:2005}.
Here,
the focus is on the nature of the transition from the Mott insulator
to the band insulator phase. We calculate the optical conductivity for
direct comparison with experimental data.  For other theoretical studies of
low-dimensional coupled strongly correlated systems see e.g.
\textcite{PotthoffNolting1998,PotthoffNolting1999,Essler:2001,Biermann2001,Biermann2002,Koga:2005}.

The rest of this paper is organized as follows: %In Sec. \ref{Section2}
First, we introduce the Hamiltonian and present the solution method
using DMFT and the Quantum Monte Carlo algorithm.
%impurity problem \cite{HirschFye1986,Hirsch1983}.
%In Sec. \ref{Section3}
We use these methods to determine the metal-to-insulator transition and
calculate the phase diagram of the model at finite temperature. Then
we consider and analyze the numerical results, in particular the
analytic continuation of the imaginary-time QMC data. We present
detailed results for the single-particle spectral function and the
optical conductivity close to the transition and analyze the behavior
of these properties close to the transition.
%Sec. \ref{Section4} contains 
Finally, we state our conclusions.
\section{Formalism}
\subsection{The model and solution method\label{Section2}}
The two-plane Hubbard model with interplane hopping $t_\perp$ is described by
the Hamiltonian
\begin{eqnarray}
H&=&-\frac{1}{\sqrt{z}}\sum_{\langle i,j\rangle\sigma\alpha}c_{i\sigma\alpha}^+
        c_{j\sigma\alpha}
-t_\perp\sum_{i\sigma\alpha}c_{i\sigma\alpha}^+c_{i\sigma,1-\alpha}
\label{Hamiltonian}\\*
&&\quad\quad\quad+ U\sum_{i\alpha}n_{i\uparrow\alpha}n_{i\downarrow\alpha}
\nonumber
\end{eqnarray}
with $c_{i\sigma\alpha}$ denoting the annihilation operator for an
electron/hole with spin component $\sigma$ on site $i$ of the
plane $\alpha=0,1$, 
and $n_{i\sigma\alpha}=c_{i\sigma\alpha}^+c_{i\sigma\alpha}$. This means
electrons can move inside the planes as well as between corresponding sites on 
the two planes. $z$ denotes the coordination number of the lattice ($z=4$ for
two dimensions), ensuring a constant band width as the coordination number
is taken towards infinity.
%/////////////////////////////////////////////////////////\\
%NOCH MEHR TEXT ?\\
%/////////////////////////////////////////////////////////\\

Using Dynamical Mean-Field theory
\cite{Metzner1989,MuellerHartmann1989,GKKR1996}, the two-plane system
is reduced to two impurities self-consistently embedded in a bath: In
order to calculate on-site (local) properties of the sites
$(i,\alpha)=(i,0), (i,1)$ (site $i$ of each of the planes), the
self-energy $\Sigma_{\alpha\alpha'} (\omega,{{\bf k}})$ is replaced by
the local self-energy $\Sigma_{\alpha\alpha'} (\omega, 0)$, leading to
a two-impurity ($i=0$, $\alpha = 0,1$) problem given by the
``effective'' action
%\begin{widetext}
\begin{subequations}
\begin{eqnarray}
&&S\left[\left(c^+_{0\sigma\alpha},c_{0\sigma\alpha}\right)_{\sigma\alpha}\right]\nonumber\\*
&&=\int{\rm d}\tau\,{\rm d}\tau'\,\sum_{\sigma\alpha\alpha'} c^+_{0\sigma\alpha}
(\tau){\cal G}_{\alpha\alpha'}(\tau,\tau')^{-1} c_{0\sigma\alpha'}(\tau')
\nonumber\\*
&&\quad+U\int{\rm d}\tau\sum_\alpha n_{0\alpha\uparrow}(\tau)
n_{0\alpha\downarrow}(\tau).
\label{EffectiveAction}
\end{eqnarray}
%\end{widetext}
The Weiss field ${\cal G}$ describes the dynamics of the site $i=0$ without the
interaction plus the rest of the lattice. ${\cal G}$ is a $2\times 2$ matrix; 
since the system is symmetric under exchange of the planes, we
 use ${\cal G}_{00}={\cal G}_{11}=:{\cal G}_0$, ${\cal G}_{01}={\cal G}_{10}=:
{\cal G}_1$; the properties of the system can be described by the 
symmetric/antisymmetric (bonding/antibonding) combinations of the two planes. 
This impurity problem is defined by the self-consistency equation
\begin{eqnarray}
{\cal G}_{\text{S/A}}\left(i\omega_n\right)^{-1}&=&\Sigma_{\text{S/A}}
        \left(i\omega_n\right)\label{SCE}\\*
&&+\tilde{D}\left(i\omega_n+\mu\mp t_\perp-\Sigma_{\text{S/A}}\left(i
\omega_n\right)\right)^{-1}\nonumber,
\end{eqnarray}
\end{subequations}
where $\tilde D(\zeta)=\int {\rm d}\varepsilon\, 
        D(\varepsilon)\left(\zeta-\varepsilon\right)^{-1}$,
 $D$ being the density of states (DOS) for a free ($U=0$) single-plane system, ${\cal
G}_{\text{S/A}}={\cal G}_0\pm{\cal G}_1$, $\Sigma_{\text{S/A}}
=\Sigma_0\pm\Sigma_1$,
and $\Sigma$ is the self-energy for the impurity problem, which can be
calculated from the effective impurity action via the impurity Green's
function (the mean-field approximation for the on-site lattice
Green's function). 
%This 
The self-sonsistency 
equation can be derived exactly following the lines given in the work by
\textcite{GKKR1996}.

Moreover, $D$ is the only place where the detailed lattice
structure enters the calculations, so the results are essentially independent
of those details.

The DMFT equations are usually solved using an iteration algorithm consisting of two
parts: By solving an impurity-like problem (\ref{EffectiveAction}), the on-site 
Green's function is determined, then, using the DMFT self-consistency equation
(\ref{SCE}), a new impurity problem is defined. This is repeated until convergence 
has apparently been reached.

We solve the two impurity problem using the Quantum Monte Carlo algorithm
developed by Hirsch and Fye \cite{HirschFye1986}. In
order to use the Monte Carlo algorithm with the DMFT effective action which is
non-local with respect to imaginary time $\tau$, the action $S$ has to be 
rewritten\cite{GKKR1996} %.
%To that end, it is necessary to express the action 
using a lattice Hamiltonian consisting of auxiliary ``bath'' orbitals, 
replacing the ``bath'' Green's function ${\cal G}$. 
% Ge"andert DH 2005-12-06

For initialization, we use a guess for the Weiss 
fields, ${\cal G}^{\text{guess}}_0(i\omega_n)$ (diagonal, i.~e., connecting one
site to itself) and ${\cal G}^{\text{guess}}_1(i\omega_n)$ (off-diagonal,
i.~e., connecting one site to the corresponding site on the other plane),
determining the Green's function of the lattice. Using the QMC algorithm, the
local imaginary-time Green's functions $G_0(\tau)$ and $G_1(\tau)$
are calculated, $G_0$ being the on-site Green's function, whereas $G_1$ is
again connecting two corresponding sites on different planes. Use of the Dyson 
equation then yields the self-energies $\Sigma_0(i\omega_n)$ 
and $\Sigma_1(i\omega_n)$.

Now, in order to use the self-consistency equation, we switch to the
symmetric/antisymmetric combinations of the two planes, so the
self-energy, the Green's function, and the Weiss field become 
diagonal $2\times 2$ matrices. Since the kinetic energy is then diagonal as well, the 
free Green's functions are the Hilbert transforms of the density of states 
for the symmetric/antisymmetric combinations of the real-space planes
without interaction:
\begin{equation}
G^0_{\text{S/A}}\left(i\omega_n\right)=\tilde D\left(i\omega_n+\mu\mp t_\perp\right).
\end{equation}
% Furthermore 
Therefore the self-energy can be easily calculated using
\begin{equation}
\Sigma_{\text{S/A}}\left(i\omega_n\right)
        =\tilde D\left(i\omega_n+\mu\mp t_\perp\right)^{-1}
        -G_{\text{S/A}}\left(i\omega_n\right)^{-1},
\end{equation} 
where $G_{\text{S/A}}=G_0\pm G_1$.
Now we can calculate the new Weiss fields for the next iteration
using the self-consistency equation (\ref{SCE}).
%\begin{eqnarray}
%{\cal G}_{\text{S/A}}\left(i\omega_n\right)^{-1}
% &=&\Sigma_{\text{S/A}}\left(i\omega_n\right)\nonumber\\*
%&&\quad+\tilde D\left(i\omega_n+\mu\mp t_\perp-\Sigma_{\text{S/A}}\left(
%       i\omega_n\right)\right)^{-1}.%\tag{(\ref{SCE})}
%\end{eqnarray}
%This equation can be derived exactly following the lines given in the work of
% A.~Georges et al. \cite{GKKR1996}.

%-----------------------------------------------------------------------------
\subsection{Optical conductivity} % subsection
Using the electron spectral densities $A_{\text{S/A}}(\omega)$, we calculate
the electron self-energy at real frequencies, $\Sigma_{S/A}(\omega)$. The
spectral function for a non-vanishing momentum is then given to be
$A_{\varepsilon}^{\text{S/A}}(\omega)=-{\text{Im}}\,G_\text{S/A}(\omega+i0,\varepsilon)/\pi
=-{\text{Im}}\left(1/\left(\omega+i0\mp t_\perp
-\varepsilon-\Sigma_{\text{S/A}}(\omega)\right)\right)/\pi$, where $\varepsilon$ is the
free-particle kinetic energy.

The optical conductivity is,
up to a constant, defined by
\begin{eqnarray}
\sigma(\nu)&=&\frac{i\sigma_0}{\nu+i\cdot 0}G_{jj}(\nu+i\cdot 0),
\end{eqnarray}
where $G_{jj}$ denotes the current-current correlation function. As a function
of the bosonic Ma\-tsu\-ba\-ra frequencies $\nu_m$, in DMFT for a hypercubic lattice, 
it is \cite{GKKR1996}
\begin{eqnarray}\label{OptCondBubble}
G_{jj}(i\nu_m)&=&\sum_{\alpha=\text{S},\text{A}}\int_{-\infty}^{\infty}{\rm d}\varepsilon\,
 D(\varepsilon)\times\\*\nonumber
&&\times\frac{1}{\beta}
\sum_{n=-\infty}^{\infty}G_{\alpha}(i\omega_n,\varepsilon) G_{\alpha}(i\omega_n+i\nu_m,\varepsilon).
\end{eqnarray}
After continuation to real frequencies $\omega$, the real
(non-dissipative) part of the optical conductivity is, up to the constant
\cite{Pruschke:1992} $\sigma_0$,
%\begin{widetext}
\begin{eqnarray}\label{OptCond} 
\text{Re}\,\sigma(\omega)&=&\sigma_0\int{\rm d}\varepsilon\,D(\varepsilon)
\sum_{\alpha=\text{S},\text{A}}\int{\rm d}\omega'\,\times\\
&&\times A^\alpha_\varepsilon(\omega')
A^\alpha_\varepsilon\left(\omega+\omega'\right)\frac{f\left(\omega'\right)
-f\left(\omega+\omega'\right)}{\omega},\nonumber
\end{eqnarray}
$f$ denoting the Fermi function, $f(\omega)=1/\left(\exp\left(\beta\omega\right)
+1\right)$, where $\beta$ denotes the inverse temperature.
Finally, the weight of the Drude peak was determined by fitting a
Lorentz curve to the central peak (the first five data points, corresponding to
$\omega<2/\beta$).% In the case of a very sharp
%peak, the Drude peak was represented by only one point at $\omega=0$;
%in this case, fitting was impossible, and the weight of the peak was
%approximated by its height times the step width of the $\omega$ grid.

%%
%
The reason for the summation over symmetric and antisymmetric planes
in eqns.  (\ref{OptCondBubble}) and (\ref{OptCond}) is the following:
since the optical conductivity is defined as a long-wavelength limit,
%since a long-wavelength limit is considered when calculating the
%optical conductivity, 
the momentum transferred by the optical
conductivity has to vanish, viz. the in-plane component as well as the
component perpendicular to the planes. The perpendicular component can
assume just two values, $0$ and $\pi/(\mbox{plane distance})$,
corresponding to symmetric and antisymmetric orbitals, respectively.
Therefore, the optical conductivity at vanishing (also perpendicular)
momentum is given by the product of Green's functions both symmetric
or both antisymmetric. --- In the limit of high dimension, inter-band
transitions do not contribute to the optical conductivity, as they may
only arise from interaction vertices. However, in that limit, the
interaction only contributes to the optical conductivity via
self-energy insertions in the single-particle Green's
function\cite{GKKR1996}, quite regardless of the detailed band-index
structure of the interaction vertex.

The replacement of the Gaussian free DOS of the hypercubic lattice by a
semicircular one (which is the exact DOS for an infinite-coordination
Bethe lattice) is an ad-hoc approximation, which may be justified by the
low weight of the Gaussian tails and their unphysicality. However,
eq. (\ref{OptCondBubble}) was derived for a hypercubic lattice with
Gaussian DOS, so, although we assume our results to be realistic, they
still are based on a substantial approximation. -- For a detailed discussion
of possible transport properties on a Bethe lattice, please refer to
\textcite{Bluemer:2003:1,Bluemer:2003:2}.

%The greater physical relevance has to be assigned to the ``symmetric'' part of 
%the optical conductivity, $\sigma_{\rm S}$. Since a long-wavelength limit is
%considered, the momentum transferred by the optical conductivity has to vanish,
%viz. the in-plane component as well as the component perpendicular to the planes.
%As the perpendicular component can assume just two values, $0$ and $2\pi/(\mbox{plane
%distance})$ corresponding to the symmetric and the antisymmetric part, the optical
%conductivity at vanishing momentum is given by $\sigma_{\rm S}$. However,
%as can be easily seen, in the case of half filling, due to the overall particle-hole 
%symmetry, the optical conductivity of the symmetric plane equals the antisymmetric one. 
%
\section{Numerical results\label{Section3}}
\subsection{Single particle density of states}
We consider the half-filled model ($n=1$) at a temperature
$T=0.025=1/\beta$, using $L=100$ time slices of $\Delta\tau=\beta/L=
0.4$. As the density of states of the free ($U=0$) uncoupled
($t_\perp=0$) lattice, we use a semicircular
$D(\varepsilon)=\sqrt{4-\varepsilon^2}/(2\pi)$, %corresponding to
which becomes exact for
electrons on a Bethe lattice \cite{GKKR1996}. This is more convenient
than a Gaussian DOS for a hypercubic lattice, because the extended unphysical
tails of the Gauss distribution render it impossible to clearly define
the metal-to-band insulator transition.

From the imaginary-time Green's functions produced by the QMC
algorithm, the corresponding spectral densities for the
symmetric/antisymmetric planes are extracted %by 
using the
Maximum-Entropy method \cite{Linden1995,JarrellGubernatis1996}. We use
a default model consisting of normalized semi-ellipses of half-width
$U/2+2$ centered at $\mp t_\perp$ for the symmetric and antisymmetric
plane, resp., plus a small flat ``background'' in order to keep the
possibility to extract features outside this area. -- Alternatively, a flat 
and a Gaussian default model were used; however, those produce unphysically 
large high-frequency tails in the spectral density and, as well, artificial 
humps at $\omega=0$ even in the non-interacting case.
% DH 2006-05-15, 2006-05-17
%----------------------- Mott-Transition Figure -------------------------------
\begin{figure}
%\begin{center}
  \includegraphics[width=8.6cm,clip]{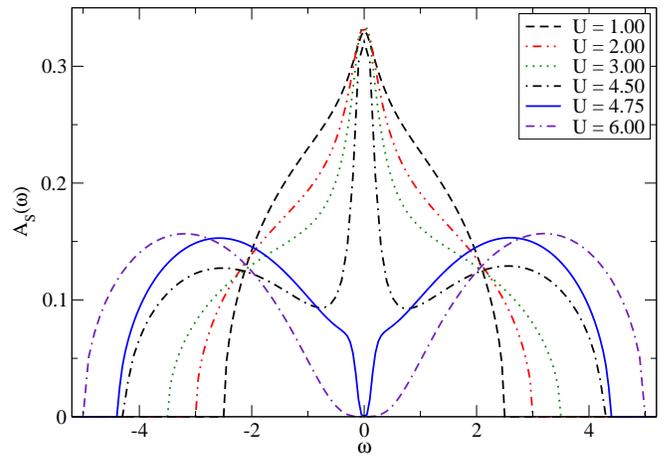}
\caption{(Color online) Mott transition. Spectral densities of the uncoupled ($t_\perp=0$) 
two-plane Hubbard model. The DOS correspond to the metallic state ($U = 1, 2, 3, 4.5$), 
and to the insulating state ($U = 4.75, 6$). The DOS at $U = 4.75$ corresponds to the 
insulating state slightly above the Mott transition. The iteration was initialized using
an insulating Green's function.}
\label{MottTransition}
%%\end{center}
\end{figure}
Figure \ref{MottTransition} shows the density of states of the
uncoupled system ($t_\perp=0$) at $U=1$, $2$, $3$, $4.5$, $4.75$, and
$6$.
%{\bf Bitte korrigieren dem Bild entsprechend.}
The DOS at $U=4.5$ has a three-peak shape characteristic for the
metallic state close to the Mott transition.
% {\bf NOCH NICHT FERTIG} 
The optical conductivity and the Drude weight yield a transition value 
$U\approx4.7$ (Fig. \ref{OC_tp0}). As the iteration was initialized using
an ``insulating'' Green's function, the transition marks the lower-U end of
the coexistence region.
The spectral density at $U=4.75$ represents the insulating state just 
after the vanishing of the quasiparticle peak. The DOS at $U=6$ displays the 
lower and upper Hubbard bands at $-U/2$ and $+U/2$, resp.

%----------------------- Band-Transition Figure -------------------------------
\begin{figure}
%\begin{center}
\includegraphics[width=8.6cm,clip]{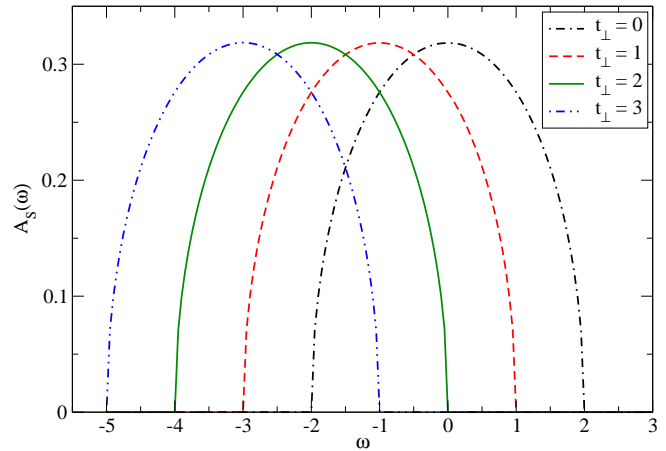}
\caption{(Color online) Band transition. Reconstructed spectral densities of the symmetric 
  plane of the free two-plane Hubbard model at $U=0$,
  $t_\perp=0,1,2,3$. The antisymmetric DOS is
  $A_{\text{A}}(\omega)=A_{\text{S}}(-\omega)$.  $A_{\text{S}}$ and
  $A_{\text{A}}$ do not overlap in the band insulating state, the band
  transition occurs at $t_\perp=2$.}
\label{BandTrans}
%\end{center}
\end{figure}
At $U=0$, the metal-to-band insulator transition was found 
at $t_\perp=2.0$ (see also the phase diagram, Fig. \ref{PhaseDiagram}). 
This is the point where the overlap of the spectral 
densities for the symmetric and the antisymmetric planes vanishes. 
%The band transition is located a little bit above the expected transition at 
%$t_\perp=2$ due to the use of a Gaussian default model in the Maximum-Entropy 
%procedure .
% Entfernt DH 2006-02-28
%%\footnote{The semicircular default model leads to the numerical
%%breakdown of the Maximum-Entropy procedure, as the Newton-Raphson Jacobian
%%matrix becomes close to singular due to the very small values of the default
%%model far away from the centre.
%% Entfernt 2005-10-05 DH}.
The spectral densities for the metallic and the band insulating phases
are given in Fig. \ref{BandTrans}. As can be seen, the spectral
densities for the symmetric/antisymmetric plane are shifted from their
$t_\perp=0$ position by exactly $\mp t_\perp$. The symmetric DOS at
$t_\perp=2$ corresponds to the
state %slightly below 
right at
the band transition. The error bars are of the order
of magnitude of the line width. 
\begin{figure}
%\begin{center}
\includegraphics[width=8.6cm,clip]{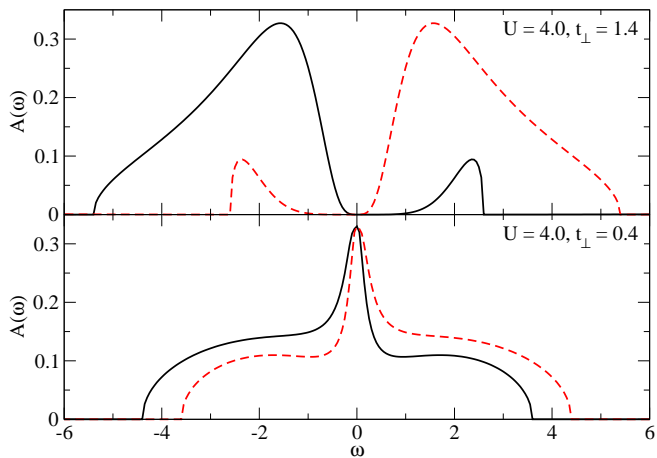}
\caption{(Color online) Reconstructed symmetric (solid line) and antisymmetric (broken line) 
  spectral densities of the two-plane Hubbard model at $U=4$ and
  $t_\perp=0.4$, $1.4$ using $L=100$ time slices.  The changes of the
  Hubbard bands due to $t_\perp$ can be clearly seen. --- Due to
  particle-hole symmetry of the two-plane system at half filling, the
  overall spectral density is symmetric.
\label{SymAntisym}}
%\end{center}
\end{figure}
For finite $U$, on increasing $t_\perp$, for the symmetric plane, the
weight of the upper Hubbard band is reduced, whereas the lower one increases,
until, at $t_\perp >2$, the upper Hubbard band has completely vanished. For
the antisymmetric plane, the upper band is increased at the expense of
the lower one. For intermediate values, this effect can be clearly
seen from Fig.~\ref{SymAntisym}.
 
Our results are compatible with earlier results for the single-plane model
found by different methods like QMC or IPT \cite{GKKR1996} or 
NRG \cite{Bulla:2001} (the upper boundary of the coexistence region is slightly
higher in our case, due to very large time slices). As discussed below,
our results are also compatible to the quantities calculated for a two-plane
model by \textcite{Moeller:1999}, although those are zero-temperature data.
%++++++++++++++++++++++++++++++++++++++++++++++++++++++++++++++++++++++++++++++
%=========================== Optical conductivity =============================
%++++++++++++++++++++++++++++++++++++++++++++++++++++++++++++++++++++++++++++++
\subsection{Optical conductivity}
%--------------------------- Optical Conductivity, tp=0 -----------------------
%{\bf Bisschen Bla-Bla ueber Peaks bei U/2 und U im metallischen Zustand, bei uns sieht man die leider nicht so deutlich. 
%Aber mal ehrlich bei Pruschke und Co sehen die Bilder nicht besser aus. Also denk dir was aus.}
\begin{figure}
%\begin{center}
\includegraphics[width=8.6cm,clip]{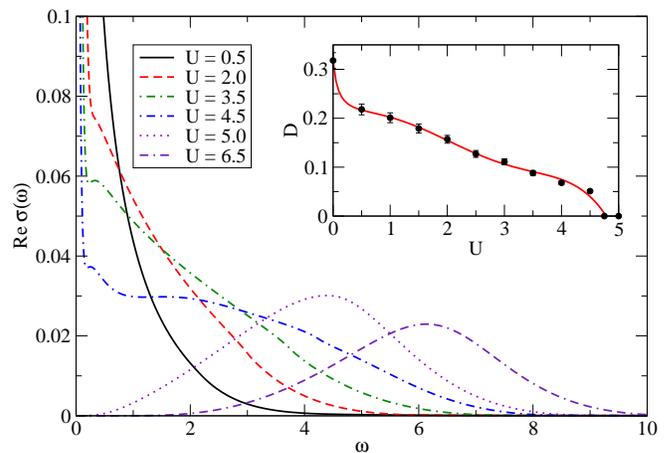}
\caption{(Color online) The evolution of the optical conductivity as a function $\omega$ at 
different $U$, $t_\perp=0$. Inset: formation of the Drude weight with 
increasing $U$. As can be seen, the Drude weight vanishes at $U\approx 4.7$.
Because the iteration was initialized using an ``insulating'' Green's function,
this is the lower end of the coexistence region.
\label{OC_tp0}
}
%\end{center}
\end{figure} 
%At first we considered the optical conductivity of the uncoupled system 
%($t_\perp=0$). Using eq. (\ref{Eq.OptCond}) followed by Maximum-Entropy data 
%analysis we found the optical conductivity for different values of $U$ (Fig.
%\ref{OC_tp0}). 
% DH 2006-02-28
At first, we consider the optical conductivity of the uncoupled system 
($t_\perp=0$). 
Using the spectral densities obtained by Maximum-Entropy,
%Using Maximum-Entropy data analysis of the current-current
%correlation function and eq. (\ref{OptCond2})
%% followed by Maximum-Entropy data analysis 
we found the optical conductivity for different values of $U$ (Fig.
\ref{OC_tp0}) using eq. (\ref{OptCond}). Our results are compatible with the
single-plane data in \textcite{Pruschke:1992}.

%The DC conductivity is equal to the weight of the Drude peak
The quasiparticle contribution to conduction is given by the weight of the Drude peak
% DH 2005-12-12
located at $\omega=0$, %at $U=0$, 
thus, an insulating system has vanishing Drude weight.
With increasing interaction parameter $U$, the Drude peak %is increasingly
%smeared. 
decreases for all values of $t_\perp$.
As well, the growth of the incoherent peak at $\omega\approx U$ is clearly 
visible. In the inset, the Drude weight %, i.~e., the DC conductivity, 
is shown as a function of $U$.
Clearly, the system becomes a Mott insulator at $U\approx 4.7$, if the iteration is initialized
with an ``insulating'' Green's function.
%
%------------------------- Optical Conductivity, U=2 --------------------------
\begin{figure}
%\begin{center}
\includegraphics[width=8.6cm,clip]{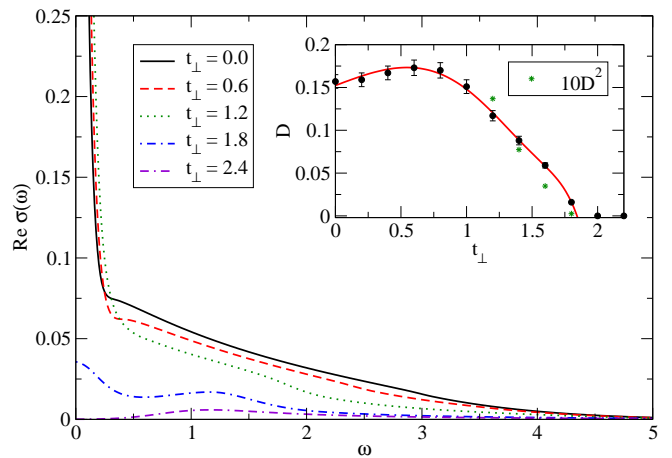}
\caption{(Color online) The evolution of the optical conductivity as a function $\omega$ at 
different $t_\perp$, $U=2$.
Inset: formation of the Drude weight with increasing $t_\perp$. As can be seen,
the Drude weight vanishes at $t_\perp\approx 1.8$, where a metal-to-insulator
transition takes place.
\label{OC_U2}
}
%\end{center}
\end{figure} 
Figure \ref{OC_U2} depicts the optical conductivity for different
$t_\perp$ at $U=2$, again consisting of the Drude peak of different weights and an
``incoherent'' part which consists of two peaks, one of them located
at $\omega\approx U$, the other one, present only in the metallic
phase, located at $\omega\approx U/2$.  However, the latter one is
usually smeared too strongly to be seen clearly\cite{Pruschke:1992},
only for low values of $t_\perp$, some traces of this peak might be
recognized.
%and the remaining weight concentrated around $\omega=U$. 
With increasing $t_\perp$, the Drude peak 
vanishes at $t_\perp\approx 1.8$ for $U=2$, indicating the transition to
a predominantly band insulating state. --- The transition value was found by
linear extrapolation of the squared Drude weight 
(see the inset of Fig. \ref{OC_U2}).
%goes to zero when $t_\perp$ grows.
%The Drude weight in the inset indicates the band transition at $t_\perp\approx
%1.65$ and $U=2$.    
%++++++++++++++++++++++++++++++++++++++++++++++++++++++++++++++++++++++++++++++
%============================ Phase Diagram ===================================
%++++++++++++++++++++++++++++++++++++++++++++++++++++++++++++++++++++++++++++++
\subsection{Phase diagram}
%---------------------------- 3D Drude ----------------------------------------
\begin{figure}
%\begin{center}
\includegraphics[width=8.6cm,clip]{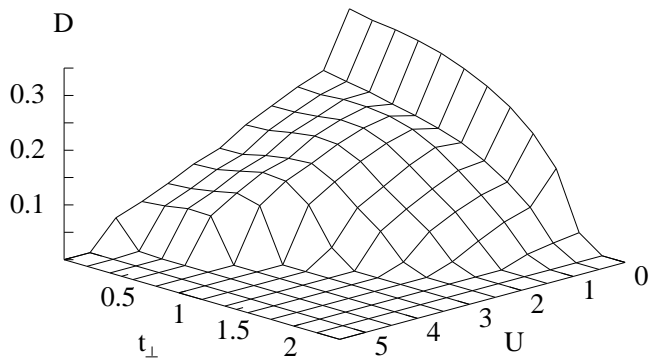}
\caption{Drude weight $D$ at temperature $T=0.025$, the iteration was initialized with an 
``insulating'' Green's function. The region $D\neq 0$ can thus
be identified as the low-$U$, low-$t_\perp$ region, as also depicted in 
Fig. \ref{PhaseDiagram}.
\label{Drude3d}} 
%\end{center}
\end{figure}
%\begin{figure}
%%\begin{center}
%\includegraphics[width=8.6cm,clip]{DrudeWU01234}
%\caption{Drude weight $D$ as a function of $t_\perp$ at temperature $T=0.05$. 
%\label{DrudeWU01234}} 
%%\end{center}
%\end{figure}

\begin{figure}
%\begin{center}
\includegraphics[width=8.6cm,clip]{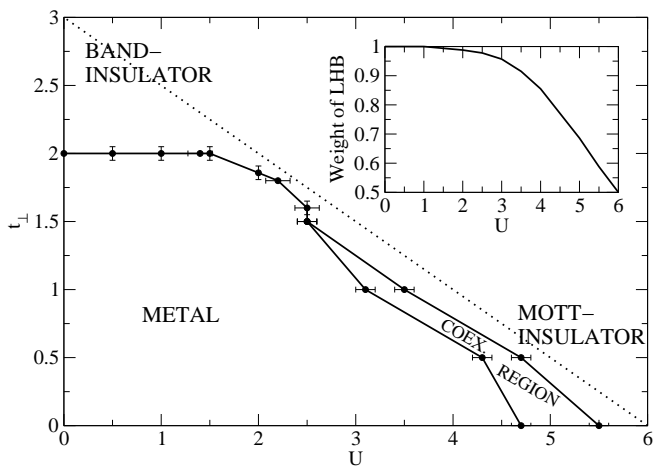}
% Best statt 1
\caption{ 
Finite-temperature ($T=0.025$) phase diagram. The lines are just a guide to the eye.
Inset: the evolution of the weight of the lower Hubbard band of 
the symmetric plane on the dotted line. --- By comparing to Fig. \ref{Drude3d},
the metallic region is recognized as the region with nonvanishing Drude
weight. 
\label{PhaseDiagram}
}
%\end{center}
\end{figure}

In Fig. \ref{Drude3d}, the Drude weights for the different values of
$(t_\perp, U)$ are shown. In order to find the metal-to-insulator transitions 
we used a linear interpolation of the quadratic value of the Drude weight,
%(see below), 
obtaining the phase diagram given in Fig. \ref{PhaseDiagram}.
% shows the evaluated phase diagram.   
%---------------------------- Phase Diagram + LHB -----------------------------
The different regions of the phase diagram could be clearly located: as expected,
there is a metallic state for low $U$ and low $t_\perp$, which is bounded by a
metal-to-band insulator transition at $t_\perp=2$. For high $U$, the system is in a Mott
insulating state; the metallic and insulating solutions are both locally stable within
a coexistence region. As discussed below, no clear separation between the Mott and the band
insulating states were found.

%The metal-to-insulator transitions could be clearly located. However, 
%considering the nature of the transition from Mott to band insulator, no
%clear evidence was found supporting a certain scenario. 
%As there is no obvious order parameter, we 
%calculated several quantities characteristic for metals or insulators.
% DH 2006-03-29

%, no clear is what happens  
%between the band insulating phase and the Mott insulating phase. We did not find an 
%appropriate quantity that we can use to answer this quastion. All we can do was the 
%calculation of the weight of the lower Hubbard band (LHB) of the symmetric plane.  
In order to get some impression of the transition between the band insulating 
and the Mott insulating phase, we calculated the weight of the lower Hubbard
band (LHB) of the symmetric plane which
\begin{figure}
%\begin{center}
\includegraphics[width=8.6cm,clip]{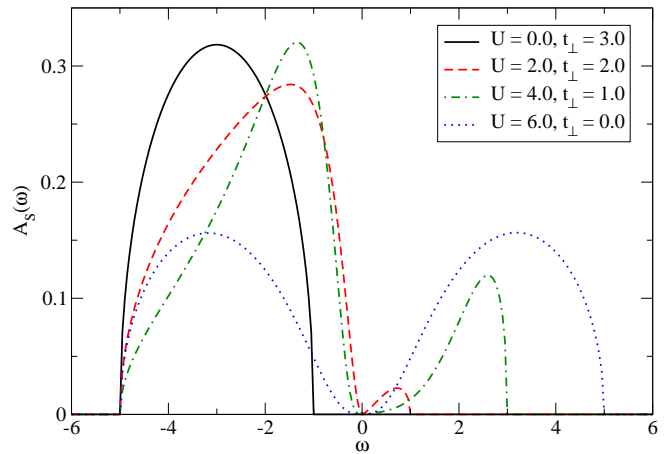}
\caption{(Color online) Selection of reconstructed symmetric spectral densities on the dotted 
line in Fig. \ref{PhaseDiagram}, at temperature $T=0.025$. --- A purely Mott
insulating state is characterized by a spectral density of the symmetric or
antisymmetric plane, resp., divided by half into a lower and an upper
Hubbard band, whereas a purely band insulating state means the symmetric
band is entirely located below $\omega=0$, and the antisymmetric band entirely
above.
\label{LineExample}
}
%\end{center}
\end{figure}
%The weight of the LHB 
is defined as $\int_{-\infty}^0{\rm d}\omega\, A_{\text{S}}(\omega)$,
%. It was calculated 
at the points given by the dotted line in Fig. \ref{PhaseDiagram}. Some of the spectral density
functions can be seen in %examples are ploted
%on the figure~
Fig. \ref{LineExample}. The evolution of the LHB weight along the dotted line 
is shown in the inset in Fig. \ref{PhaseDiagram}. For half filling, the weight 
of the LHB for a  purely band insulating phase is unity, for a purely Mott 
insulating phase, it is close to $0.5$.
%lies between $0.5$ and $1$. 
As can be seen the weight of the LHB does neither show some distinct kink nor 
vanish from a well-defined point. 
Thus, this quantity does not yield any evidence for a phase transition between
the Mott-Hubbard and the band insulating phase.

\begin{figure}
\includegraphics[width=8.6cm,clip]{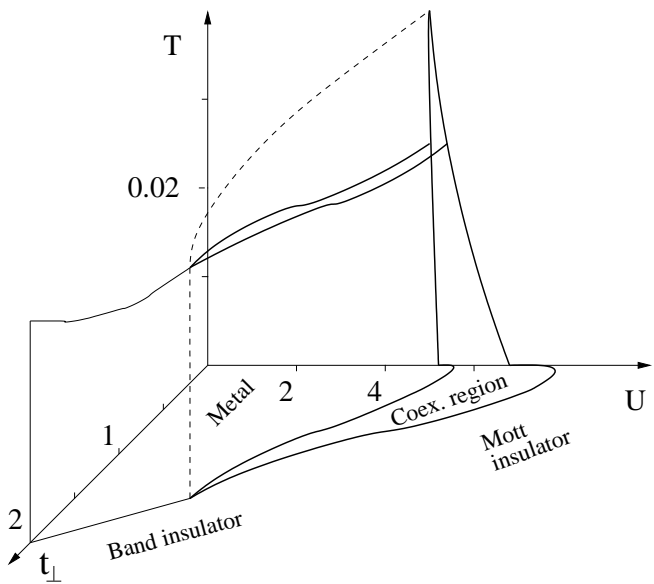}
\caption{\label{Coexistence3D}Three-dimensional phase diagram of the two-plane
Hubbard model, composed of the data in \textcite{Moeller:1999} ($T=0$), \textcite{GKKR1996}
($t_\perp=0$), and this work. The dashed line indicates the shape of the coexistence
region suggested by the data, which are indicated by the full lines; thin lines indicate the
metal-to-insulator transitions.}
\end{figure}

We find a phase diagram which is clearly compatible to the zero-temperature
phase diagram in \textcite{Moeller:1999}, keeping in mind that
$U_{\text{Moeller}}=U/2$ and $t_{ab\text{Moeller}}=t_\perp/2$. However, some
differences ought to be noticed: the %Mott transition
coexistence region %for the uncoupled planes
is found at a lower $U$ value due to the finite temperature (see, for comparison,
the phase diagram in \textcite{GKKR1996}, where the same scale of $U$ is used as
by \textcite{Moeller:1999});
%we also see no hysteresis because our temperature is higher than the critical
%temperature. 
the coexistence region has become smaller as well. A coexistence region thus clearly
exists at a temperature of $T=0.025$; in contrast, at $T=0.05$, no coexistence region was
found any more. --- This behavior suggests that the critical temperature of the
Mott transition decreases as $t_\perp$ is increasing, see the sketch in Fig. \ref{Coexistence3D}.
 
The other clear difference is the slope of the transition line at low interaction
$U$, close to the band transition. We find, for low $U$, the transition line
to be at almost constant $t_\perp$, whereas Moeller et al. find a clear dependence
on $t_\perp$. We assume this is due to the temperature: as a finite temperature
always smoothens a metal-to-band insulator transition, a small interaction driving
the system to an insulating state can be compensated by thermal fluctuations.
% DH 2005-12-15
% stimmt das, oder ist das dummes Blabla???

%Additionally, the re-entrance behavior found by \textcite{Moeller:1999} for
%low $t_\perp$ could not be seen in our work. As no $t_\perp$ values in
%the region $0<t_\perp < 0.5$, were studied, this does not imply there is no
%re-entrance behavior on a smaller scale.
%%DH 2006-05-12

The re-entrance behavior seen in the IPT\cite{Moeller:1999} cannot be resolved
accurately in our calculation. The general shape of the phase boundary however
suggests that a re-entrance behavior does not exist at the temperature
considered. To resolve this issue definitely, lower temperatures have to be
considered which are inaccessible to the Hirsch-Fye algorithm for the
large-$U$ case.
% HM 2006-05-17

The comparatively\cite{GKKR1996} high upper bound of the coexistence region 
may be due to the non-negligible Trotter error in this region.
The insulating solution will be much less sensitive to an increase in the 
$(U \Delta\tau)^2$ term neglected by the Trotter decomposition of the path 
integral than the metallic solution, which is why the upper bound shifts due 
to the truncation error, but not so much the lower bound, which is also at 
a lower $U$ value, so the second-order $U$ term neglected by the time 
discretization is smaller.

\section{Summary}
In summary, we have calculated the spectral densities, the optical conductivities,
and the Drude weights of a two-plane Hubbard
model at low temperature for different values of the inter-plane coupling. We
have located the different metal-to-insulator transitions; however, no clear
transition between the Mott insulating phase and the band insulating phase
could be found; as well, the corresponding spectral weights show a continuous
behavior. This observation is consistent with the assumption that there
exists only a crossover between those two insulating phases, but no clear phase
transition.
The phase diagram is slightly different from the one found in
\textcite{Moeller:1999}, which comes as no surprise as we are considering a finite
temperature % possibly above a coexistence region.
%Therefore, the existence of such a coexistence region at lower but finite
%temperature still remains unclear.
shifting transition values and decreasing the coexistence region.

We have discussed in detail spectral properties like, e.~g., the optical
conductivity, which is of some use to experimentalists. Also, the use of
a Quantum Monte Carlo method means a serious technical improvement with
respect to earlier studies, as it is a numerically exact method without the 
use of uncontrolled approximations.

%%The evolution of the system towards a higher spatial dimensionality is,
%%in this context, given by changing the inter-plane coupling parameter $t_\perp$
%%from zero to higher values. 
%In this context, changing the value of the inter-plane hopping parameter
%from zero to small values $t_\perp$ re-introduces some aspect of finite
%dimensionality and mimicks the evolution of a finite-dimensional system towards
%a higher dimensionality.
%%
%We found no immediate change of any quantity. %;
%%only at $t_\perp>0.4$, the Mott transition is slightly affected by the 
%%coupling of the planes. The ``critical'' exponent of the Drude weight
%%smoothly changes when increasing $t_\perp$ but the quality of the fitting
%%parameters is not sufficient to support any statement about a change of
%%the Mott transition in this case. 
%Thus, we also have no hint to a change of 
%nature of the Mott transition on increasing dimensionality.

% ge''andert und alles auskommentiert, DH 2006-05-12

Even though the use of a Quantum  MC algorithm
\cite{HirschFye1986} means a technical advantage, %only comparatively high 
%temperatures could be 
the behavior of the system for very low temperatures could not be
% DH 2006-03-29
considered in this work. We plan to investigate lower 
temperatures using a very recently developed continuous-time Quantum Monte 
Carlo algorithm \cite{Lichtenstein:2004,Lichtenstein:2005}, yielding the phase
diagram at much lower temperatures 
% without the bias necessarily introduced by the use of IPT methods 
and clarifying the evolution towards
zero temperature. % as well as the emergence of a coexistence region and the
%development of a first-order transition. 
%First successful attempts to apply 
%the continuous-time QMC algorithm to two-impurity problems already exist 
%\cite{Savkin:2005}.
%
\begin{acknowledgments}
We are thankful for the hospitality of the Aspen Centre for Physics (H.M.),
Rutgers University, and Columbia University. Fruitful discussions with 
A.~Georges, M.~Imada, G.~Kotliar, W.~Krauth, A.~I.~Lichtenstein, A.~J.~Millis, 
and O.~Parcollet are gratefully acknowdeged.

This work was supported by grants SFB 608 and SPP 1073 of the Deutsche
Forschungsgemeinschaft and a grant by the DAAD (German Academic Exchange 
service) (A.F.). 
\end{acknowledgments}
%

%

%

%

%\bibliographystyle{apsrev} %alpha, acm,myalpha,JHEP,myplain2
%\bibliography{TwoPlanePaper}

%\end{landscape}
\end{document}